\documentclass[twocolumn]{IEEEtran}
\usepackage[usenames,dvipsnames]{color}
\usepackage{subfigure}
\usepackage{graphicx}
\usepackage{amsmath}
\usepackage{color}
\usepackage{multicol}
\usepackage{amssymb}

\usepackage{amsfonts}%
\usepackage{verbatim}%
\usepackage{enumerate}%
\usepackage{psfrag}
\usepackage{epsfig}
\usepackage{multicol}
\usepackage{setspace}
\usepackage{dsfont}

\usepackage{algorithm}
\usepackage{algorithmic}
\usepackage{cite}
\usepackage{float}

\begin{document}

\title{Cooperative Cognitive Relaying with Ordered Cognitive Multiple Access}
\author{ Ahmed El Shafie$^\dagger$, Ahmed Sultan$^*$\\
\small \begin{tabular}{c}
$^\dagger$Wireless Intelligent Networks Center (WINC), Nile University, Giza, Egypt. \\
$^*$Department of Electrical Engineering, Alexandria University, Alexandria, Egypt. \\
\end{tabular}
}
\date{}
\maketitle
\begin{abstract}
We investigate a cognitive radio system with two secondary users who can cooperate with the primary user in relaying its packets to the primary receiver. In addition to its own queue, each secondary user has a queue to keep the primary packets that are not received correctly by the primary receiver. The secondary users accept the unreceived primary packets with a certain probability and transmit randomly from either of their queues if both are nonempty. These probabilities are optimized to expand the maximum stable throughput region of the system. Moreover, we suggest a secondary multiple access scheme in which one secondary user senses the channel for $\tau$ seconds from the beginning of the time slot and transmits if the channel is found to be free. The other secondary user senses the channel over the period $[0,2\tau]$ to detect the possible activity of the primary user and the first-ranked secondary user. It transmits, if possible, starting after $2\tau$ seconds from the beginning of the time slot. It compensates for the delayed transmission by increasing its transmission rate so that it still transmits one packet during the time slot. We show the potential advantage of this ordered system over the conventional random access system. We also show the benefit of cooperation in enhancing the network's throughput.
\end{abstract}
\section{Introduction}

In a wireless communication network with many source-destination pairs,
cooperative transmission by relay nodes has the potential
to improve the overall network performance. Cognitive relaying is an integration between cognitive radios and cooperative transmission that could considerably improve the performance tradeoffs for all users. The problem of cognitive relaying has been considered in many papers, for instance \cite{cogrelay1,simeone,veryclose,close,krikidis2010stability,khattab,erph,cogrelay}. Note that cooperation requires changes in the primary network. Although this may seem to go against one conception of the secondary network operating with little or no modification at all at the primary terminals, the proposed cooperative cognitive schemes show clearly that the primary networks can benefit from secondary cooperation. The realized benefits may provide a strong incentive for implementing some changes in the primary network. In addition, there can be additional monetary compensation from the secondary network, but this issue is outside the scope of this paper. 

In \cite{cogrelay1,simeone}, the main idea of a cooperative cognitive user was introduced, where the secondary user is used as a relay for the undelivered packets of the higher-priority primary user (PU). To accomplish this feat, the secondary user has an additional relaying queue to store primary packets. An acceptance factor controls the fraction of undelivered primary packets that gets accepted into the relaying queue at the secondary terminal. 

Two secondary and one primary nodes are investigated in \cite{veryclose} under a random access scheme. All nodes communicate with the same access point. Priority in secondary transmission is given to the relaying queue holding the primary packets. In addition to conventional spectrum sensing, opportunistic sensing is employed where the secondary users sense the channel at different times depending on the quality of the channel to the access point. Multiple secondary users are considered in \cite{close} together with one PU. The secondary users construct a cluster with a common relaying queue  in order to relay the undelivered packets of the PU. The packet is added to the common relaying queue, which is accessible from all the nodes of the cluster. In \cite{erph}, the authors characterize the stable-throughput region in a two-user cognitive shared channel with multipacket reception (MPR) capability added to the physical layer of the nodes. 

In this paper, we analyze a system with one PU and multiple secondary users operating in a time-slotted fashion. We present here results for two cognitive users and leave the general case for an extended version of this work. In contrast with \cite{close}, we assume that {\bf every} secondary user has an additional queue that can be used to store primary packets and help in relaying them to the primary receiver. When a transmission opportunity is available, a cognitive user randomly selects to transmit from its queue or the relaying queue. We propose an ordered access protocol where the secondary users are ordered in terms of accessing the channel. The users are also ordered in terms of their attempts to decode the received primary packets. The ordering is probabilistic, unlike \cite{veryclose} where a deterministic sensing order is proposed. The probability of each possible permutation of users indicates the fraction of time slots in which the permutation is employed. The ordering probabilities are optimized given all the system parameters and do not just depend on the quality of the channel to the receiver as in \cite{veryclose}.

We summarize our contributions as follows. For two secondary users, we characterize outer and inner bounds for the maximum stable throughput of an ordered access/decoding scheme. We obtain the optimal queue selection probabilities, acceptance fractions, and probabilistic orders to achieve the bounds. We compare our scheme with conventional random access and show its potential benefit. Note that in \cite{veryclose} only inner bounds are provided and the authors do not consider the impact of delayed access on the outage probability.

The rest of the paper is organized as follows. We provide a detailed account of the system model in Section \ref{systemmodel}. We introduce the ordered access scheme and its stability analysis in Section \ref{problemformulation}. The random access system is discussed in Section \ref{randomaccess}. We provide some numerical results in Section \ref{numericalresults} and conclude the paper in Section \ref{conclusion}.
\vspace{-0.2 cm}
\section{SYSTEM MODEL} \label{systemmodel}
\begin{figure}[t]
  \includegraphics[width=0.89\columnwidth]{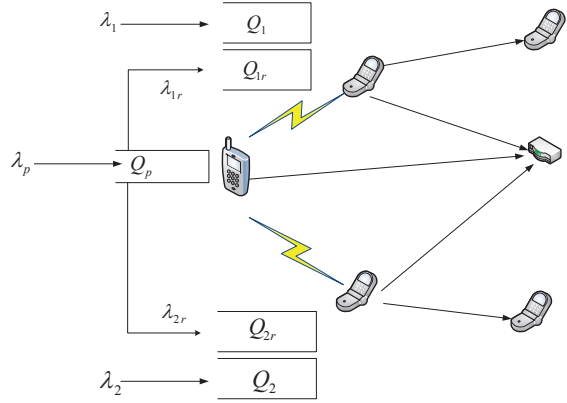}\\
  \caption{Primary and secondary links and queues.}\label{proposal}
  \vspace{-0.5 cm}
\end{figure}

We consider a time-slotted synchronized cognitive relaying system, as depicted in
Fig.\ \ref{proposal}, with one primary and two secondary users: $s_1$ and $s_2$. All nodes have buffers with infinite capacity. The PU has one queue, $Q_{\rm p}$, whereas secondary user $s_j$, $j \in \{1,2\}$, has two queues: $Q_j$ to store its own packets, and $Q_{j{\rm r}}$ to possibly store some of the primary packets until they are relayed to the primary receiver. The arrivals at queues $Q_{\rm p}$, $Q_1$ and $Q_2$ are independent Bernoulli random variables with means $\lambda_{\rm p}$, $\lambda_1$ and $\lambda_2$, respectively. Spectrum sensing is assumed to be perfect in this work.

We assume that $s_1$ and $s_2$ are ordered in terms of sensing and accessing the primary channel such that the probability of $s_1$ being ranked first is $\epsilon$. This means that in a large number of time slots, user $s_1$ attempts to access the channel first in a fraction $\epsilon$ of them. Moreover, $s_1$ and $s_2$ are ordered in terms of decoding a correctly received primary packet near the end of the time slot such that the probability of $s_1$ being ranked first is $\rho$. This means that over a large number of time slots, user $s_1$ attempts to decode the primary packet first in a fraction $\rho$ of them.

The PU transmits the packet at the head of its queue at the beginning of the time slot. The secondary terminals operate their receivers/spectrum sensors from the beginning of the time slots. The first-ranked secondary transmitter, which is $s_1$ with probability $\epsilon$ or $s_2$ with probability $1-\epsilon$, utilizes the received signal over the first $\tau$ seconds of the slot to determine the state of primary activity. If the channel is sensed to be free and the secondary user has a packet to send, it will transmit over the remaining slot duration. Note that $\tau$ should be large enough to validate the perfect sensing assumption.

The second-ranked user gathers samples from primary transmission over the first $\tau$ seconds of the time slot to detect the possible activity of the PU, and over the interval $[\tau,2\tau]$ to detect the activity of the first-ranked secondary user. If the channel is sensed to be free and if it has a packet to send, the second-ranked user switches to the transmission mode starting from time $2\tau$. The user still sends one full packet by adapting its transmission rate at the price of an increased probability of link outage as shown later. 

If the channel is sensed to be busy due to the primary activity, the secondary users continue receiving the primary transmission till near the end of the time slot. If the primary receiver acknowledges the correct reception of the primary transmitted packet by sending an acknowledgment (ACK) message, the secondary terminals discard what they have received from the PU. If the primary receiver declares its failure to decode the primary packet correctly by generating a negative-acknowledgment (NACK) message, the secondary terminals attempt to decode the primary packet. We assume that the overhead for transmitting the ACK and NACK messages is very small compared to packet sizes. In addition, we assume the perfect decoding of the ACK and NACK messages at the primary and secondary users.

The first-ranked terminal in terms of decoding the primary packet is $s_1$ with probability $\rho$ or $s_2$ with probability $1-\rho$. If it decodes the primary packet correctly, it accepts it with a certain probability into a special relaying queue dedicated to keeping the primary packets. It then issues an ACK signal so that the packet is removed from the queues of the PU and the other secondary user. If the decoding fails or the packet is not accepted, the second-ranked secondary user attempts to decode it. If decoding succeeds, it accepts the packet with a certain probability, else the packet is discarded. If no secondary terminal accepts the packet, it is retained in the primary queue. The medium access control (MAC) operation can be summarized as follows:
\vspace{-0.1 cm}
\begin{itemize}
\item At the beginning of a time slot, the primary node transmits the packet at the head of its queue. 
\item If the first-ranked user detects a free channel, it randomly chooses one of its queues to transmit a packet from if both are nonempty. The second-ranked secondary user does the same after detecting both the PU and the first-ranked cognitive user to be idle.
\item If the PU is active and the transmitted packet is received correctly by the primary receiver, an ACK message is fed back from the receiver. The packet is then dropped from $Q_{\rm p}$. The secondary nodes also discard what they have received from primary transmission.
\item If the primary packet is not received correctly, a NACK message is fed back from the primary receiver. Based on their primary packet decoding ranking, the first-ranked secondary user tries to decode the primary packet. If the packet is decoded correctly, it is accepted with a certain probability and an ACK message is transmitted, thereby inducing the primary transmitter to drop the packet. \footnote{We assume that the conventional ACK/NACK protocol of the primary network is modified such that the primary transmitter is notified about successful reception by either its respective receiver or any secondary transmitter. As mentioned in the Introduction, the higher throughput gains by the primary network may provide an incentive for implementing such protocol modifications.} If the first-ranked cognitive user fails to decode the primary packet or does not accept it, the second-ranked user tries to do so. This user issues an ACK signal if it decodes the packet successfully and decides to accept it. If no secondary user accepts the packet, it is kept in $Q_{\rm p}$ for retransmission.
\end{itemize}

\vspace{-0.1 cm}
\section{Problem Formulation}\label{problemformulation}
\vspace{-0.1 cm}
We focus in this paper on the stability of system queues, which is a fundamental performance measure of a communication network. Stability can be defined as follows. Denote by $Q^{t}$ the length of queue $Q$ at the beginning of time slot $t$. Queue $Q$ is said to be stable if \cite{sadek}
\begin{equation}\label{stabilityeqn}
    \lim_{x \rightarrow \infty  } \lim_{t \rightarrow \infty  } {\rm Pr}\{Q^t<x\}=1 
\end{equation}
\noindent We can apply Loynes' theorem to check for queue stability \cite{loynes1962stability,sadek}. This theorem states that if the arrival and service processes of a queue are strictly stationary, and the average service rate is greater than the arrival rate of the queue, then the queue is stable, otherwise the queue is unstable. A multiqueue system is stable when every queue in the system is stable.

Let the cognitive node $s_j$, $j \in \{1,2\}$, accept a correctly received primary packet with probability $f_j$. Since the probability that $s_1$ is ranked first in terms of decoding the primary packet is $\rho$, the primary service rate can be expressed as:
\begin{equation}
\begin{split}
\mu_{\rm p}&=\overline{P}_{\rm out,p}+P_{\rm out,p}\Bigg[\rho\bigg(\overline{P}_{{\rm out,p}1}f_1+\overline{\overline{P}_{{\rm out,p}1}f_1}\overline{P}_{{\rm out,p}2}f_2\bigg)\\
&+\overline{\rho}\bigg(\overline{P}_{{\rm out,p}2}f_2+\overline{\overline{P}_{{\rm out,p}2}f_2}\overline{P}_{{\rm out,p}1}f_1\bigg)\Bigg]
\end{split}
\end{equation}
\noindent where $\overline{x}=1-x$, $P_{\rm out,p}$ is the probability that the primary channel is in outage, ${P}_{{\rm out,p}1}$ and ${P}_{{\rm out,p}2}$ are the probabilities of the channels between the PU and $s_1$ and $s_2$, respectively, being in outage. Note that the primary packet is accepted by the second-ranked cognitive user if it is decoded correctly by that user and if it is not admitted into the relaying queue of the first-ranked user. Note also the benefit that accrues to the PU as a result of cooperation. Without cooperation, the primary service rate is $\overline{P}_{\rm out,p}$. When the secondary users help in relaying the primary packets, the service rate increases unless $f_1=f_2=0$, which is the noncooperative case. 
\noindent The service rate $\mu_{\rm p}$ can be written as
\begin{equation}
\begin{split}
\mu_{\rm p}&=\overline{P}_{\rm out,p}
\\&+P_{\rm out,p}\Bigg[\overline{P}_{{\rm out,p}1}f_1+\overline{P}_{{\rm out,p}2}f_2-\overline{P}_{{\rm out,p}1}f_1\overline{P}_{{\rm out,p}2}f_2\Bigg]
\end{split}
\end{equation}
\noindent The probability that the primary queue is empty is
\begin{equation}
\pi_{p_e}=1-\frac{\lambda_{\rm p}}{\mu_{\rm p}}
\end{equation}
\noindent 

\noindent The arrival rates to the queues $Q_{1{\rm r}}$ and $Q_{2{\rm r}}$ are 
\begin{equation}
\begin{split}
\lambda_{1{\rm r}}=\overline{\pi_{p_e}}P_{\rm out,p}\overline{P}_{{\rm out,p}1}f_1\left(\rho+\overline{\rho}\;\overline{\overline{P}_{{\rm out,p}2}f_2}\right)
\end{split}
\end{equation}
\begin{equation}
\begin{split}
\lambda_{2{\rm r}}=\overline{\pi_{p_e}}P_{\rm out,p}\overline{P}_{{\rm out,p}2}f_2\left(\overline{\rho}+\rho\;\overline{\overline{P}_{{\rm out,p}1}f_1}\right)
\end{split}
\end{equation}

Recall that the two secondary users are ranked in terms of their order of transmission such that the probability of $s_1$ being first is $\epsilon$. If secondary user $s_1$ is to transmit, it does so from $Q_1$ or $Q_{1{\rm r}}$. If both queues are nonempty, it transmits from $Q_{1}$ with probability $p_1$. Similarly, if secondary user $s_2$ is to transmit, it does so from $Q_2$ or $Q_{2{\rm r}}$. If both queues are nonempty, it transmits from $Q_{2}$ with probability $p_2$.
The probability of {\bf correct} packet reception by secondary terminal $j$ if it starts transmission after $i\tau$ seconds, $i=1,2,..$, is given by \cite{ShafieSultan}:
\begin{equation}
\overline{P}_{{\rm out},ij}=a\exp\Bigg(-b\;{2^{\frac{c}{\left(1-i\frac{\tau}{T}\right)}}}\Bigg)
\label{correc}
\end{equation}
\noindent where $a$, $b$, and $c$ are constants that depend on the bandwidth of transmission, the packet size in bits, the slot duration, the transmit power and the average secondary channel gain. Note that this probability decreases as $i$ increases.

Based on the aforementioned probabilities, the service rates of the secondary queues can be written as  
\begin{equation}
\begin{split}
\mu_1&=\pi_{p_e}\overline{P}_{{\rm out},11}\bigg[\epsilon+\overline{\epsilon} \delta_1{\rm Pr}\{Q_2=0,Q_{2{\rm r}}=0\}\bigg]\\&\cdot\bigg[p_1{\rm Pr}\{Q_{1{\rm r}}\neq 0\}+{\rm Pr}\{Q_{1{\rm r}}= 0\}\bigg]
\end{split}
\end{equation}
\begin{equation}
\begin{split}
\mu_2&=\pi_{p_e}\overline{P}_{{\rm out},12}\bigg[\overline{\epsilon}+\epsilon \delta_2{\rm Pr}\{Q_1=0,Q_{1{\rm r}}=0\}\bigg]\\&\cdot\bigg[p_2{\rm Pr}\{Q_{2{\rm r}}\neq 0\}+{\rm Pr}\{Q_{2{\rm r}}= 0\}\bigg]
\end{split}
\end{equation}
\noindent where $\delta_1=\frac{\overline{P}_{{\rm out},21}}{\overline{P}_{{\rm out},11}}$ and $\delta_2=\frac{\overline{P}_{{\rm out},22}}{\overline{P}_{{\rm out},12}}$.

The service rates of the relaying queues at the secondary nodes are
\begin{equation}
\begin{split}
\mu_{1{\rm r}}&=\pi_{p_e}\overline{P}^{\left({\rm P}\right)}_{{\rm out},11}\bigg[\epsilon+\overline{\epsilon}\; \delta^{\left({\rm P}\right)}_1{\rm Pr}\{Q_2=0,Q_{2{\rm r}}=0\}\bigg]\\&\cdot\bigg[\overline{p_1}{\rm Pr}\{Q_{1}\neq 0\}+{\rm Pr}\{Q_{1}= 0\}\bigg]
\end{split}
\end{equation}
\begin{equation}
\begin{split}
\mu_{2{\rm r}}&=\pi_{p_e}\overline{P}^{\left({\rm P}\right)}_{{\rm out},12}\bigg[\overline{\epsilon}+\epsilon\; \delta^{\left({\rm P}\right)}_1{\rm Pr}\{Q_1=0,Q_{1{\rm r}}=0\}\bigg]\\&\cdot\bigg[\overline{p_2}{\rm Pr}\{Q_{2}\neq 0\}+{\rm Pr}\{Q_{2}= 0\}\bigg]
\end{split}
\end{equation}
\noindent where $\overline{P}_{{\rm out,}ij}^{\left({\rm P}\right)}$ is the probability that the channel between secondary user $j$ and the primary receiver is not in outage when the secondary node transmits starting at the instant $i\tau$ relative to the beginning of the time slot. Probability $\overline{P}_{{\rm out,}ij}^{\left({\rm P}\right)}$ has a form similar to (\ref{correc}) with the channel parameters relevant to the secondary transmission to the primary receiver. Parameters  $\delta_1^{\left({\rm P}\right)}=\frac{\overline{P}^{\left({\rm P}\right)}_{{\rm out},21}}{\overline{P}^{\left({\rm P}\right)}_{{\rm out},11}}$ and $\delta_2^{\left({\rm P}\right)}=\frac{\overline{P}^{\left({\rm P}\right)}_{{\rm out},22}}{\overline{P}^{\left({\rm P}\right)}_{{\rm out},12}}$.

Our main objective in this paper is to characterize the {\bf stability region} defined as the set of secondary arrival rate pairs $\left(\lambda_1,\lambda_2\right)$ such that the system queues are stable. Since $Q_{\rm p}$ does not interact with the other queues because spectrum sensing is assumed to be perfect, its stability is independent of secondary operation and is guaranteed when $\lambda_{\rm p}< \mu_p$. On the other hand, since the queues of $s_1$ and $s_2$ are interacting and their exact analysis is intractable, we provide inner and outer bounds on the stability region.

\vspace{-0.4 cm}
\subsection{Inner Bound}

We derive the inner bound by utilizing the concept of dominant systems and by investigating the case in which the secondary nodes do not cooperate with the PU. A dominant system is constructed such that the queues in the dominant system are never less than those of the original system, provided that all queues are initialized identically \cite{rao1988stability}. Consequently, stability conditions of a dominant system are {\bf sufficient} for the stability of the original system \footnote{In a few cases, the dominant system method produces an exact result for the stability region, e.g., \cite{rao1988stability}.}.


\subsubsection{First dominant system}
In the first dominant system, the secondary users transmit dummy packets if the queues $Q_1$ and $Q_{2{\rm r}}$ are empty. Queues $Q_2$ and $Q_{1{\rm r}}$ behave as they would in the original system in the sense that no transmission from them is possible when they are empty. Under such operational assumptions, ${\rm Pr}\{Q_1=0\}={\rm Pr}\{Q_{2{\rm r}}=0\}=0$. Therefore, the service rates $\mu_{1{\rm r}}$ and $\mu_2$ can be expressed as: 
\begin{equation}
\mu_{1{\rm r}}=\pi_{p_e}\overline{P}^{\left({\rm P}\right)}_{{\rm out},11}\;\epsilon\;\overline{p_1}
\end{equation}
\begin{equation}
\mu_2=\pi_{p_e}\overline{P}_{{\rm out},12}\;\overline{\epsilon}\;p_2
\end{equation}
\noindent The probabilities ${\rm Pr}\{Q_2=0\}$ and ${\rm Pr}\{Q_{1{\rm r}}=0\}$ are given by
\begin{equation}
{\rm Pr}\{Q_2=0\}=1-\frac{\lambda_2}{\mu_2},\,\,\,{\rm Pr}\{Q_{1{\rm r}}=0\}=1-\frac{\lambda_{1{\rm r}}}{\mu_{1{\rm r}}}
\end{equation}
\noindent Therefore,
\begin{equation}
\mu_1=\pi_{p_e}\overline{P}_{{\rm out},11}\;\epsilon\;\bigg[p_1\frac{\lambda_{1{\rm r}}}{\mu_{1{\rm r}}}+1-\frac{\lambda_{1{\rm r}}}{\mu_{1{\rm r}}}\bigg]
\end{equation}
\begin{equation}
\begin{split}
\mu_{2{\rm r}}=\pi_{p_e}\overline{P}^{\left({\rm P}\right)}_{{\rm out},12}\overline{\epsilon}\bigg[\overline{p_2} \frac{\lambda_2}{\mu_2}+1-\frac{\lambda_2}{\mu_2}\bigg]
\end{split}
\end{equation}
\noindent The stability region based on the first dominant system is given by the closure of the rate pairs $(\lambda_1,\lambda_2)$ constrained by the stability of the queues as the system parameters vary over all their possible values. One method to characterize this closure is to solve a constrained optimization problem to find the maximum feasible $\lambda_1$ corresponding to each feasible $\lambda_2$ over all the possible values of $\epsilon,p_1,p_2,f_1,f_2$ and $\rho$ \cite{erph,sadek}. The optimization problem is given by:
\begin{equation}
\begin{split}
&\max_{\epsilon,p_1,p_2,f_1,f_2,\rho} \lambda_1=\pi_{p_e}\overline{P}_{{\rm out},11}\;\epsilon\;\bigg[p_1\frac{\lambda_{1{\rm r}}}{\mu_{1{\rm r}}}+1-\frac{\lambda_{1{\rm r}}}{\mu_{1{\rm r}}}\bigg]\\
&\,\,\,\,\,\,\,\,{\rm s.t.}\,\,\,\,0\le \epsilon,p_{1}, p_{2},f_1,f_{2},\rho \le 1\\
&\,\,\,\,\,\,\, \,\,\,\,\,\,\,\,\,\,\,\,\lambda_{\rm p} \leq \mu_{\rm p},\, \lambda_{1{\rm r}}\leq \mu_{1{\rm r}},\,\lambda_{2{\rm r}}\leq \mu_{2{\rm r}},\,\lambda_2\leq \mu_2
\end{split}
\end{equation}
\noindent This optimization problem and the others presented in this work are solved numerically\footnote{Specifically, we use Matlab's fmincon. Since the problems are nonconvex, fmincon produces a locally optimum solution. To enhance the solution and increase the likelihood of obtaining the global optimum, the program can be run many times with different initializations of the optimization variables.}.

\subsubsection{Second dominant system}
In the second dominant system, dummy packets are transmitted from $Q_2$ and $Q_{1{\rm r}}$ when these queues are empty. Consequently, ${\rm Pr}\{Q_2=0\}={\rm Pr}\{Q_{1{\rm r}}=0\}=0$. The operation of $Q_1$ and $Q_{2{\rm r}}$ is retained as in the original system.  Similar to the first dominant system, the service rates of the queues are as follows:
\begin{equation}
\mu_1=\pi_{p_e}\overline{P}_{{\rm out},11}\;\epsilon\; p_1
\end{equation}
\begin{equation}
\begin{split}
\mu_{1{\rm r}}=\pi_{p_e}\overline{P}^{\left({\rm P}\right)}_{{\rm out},11}\;\epsilon\;\bigg[\overline{p_1}\frac{\lambda_1}{\mu_1}+1-\frac{\lambda_1}{\mu_1}\bigg]
\end{split}
\end{equation}
\begin{equation}
\begin{split}
\mu_{2{\rm r}}=\pi_{p_e}\overline{P}^{\left({\rm P}\right)}_{{\rm out},12}\;\overline{\epsilon}\;\overline{p_2}
\end{split}
\end{equation}
\begin{equation}
\mu_2=\pi_{p_e}\overline{P}_{{\rm out},12}\;\overline{\epsilon}\;\bigg[p_2\frac{\lambda_{2{\rm r}}}{\mu_{2{\rm r}}}+1-\frac{\lambda_{2{\rm r}}}{\mu_{2{\rm r}}}\bigg]
\end{equation}
\noindent The relevant optimization problem is given by:
\begin{equation}
\begin{split}
&\max_{\epsilon,p_1,p_2,f_1,f_2,\rho} \lambda_2=\pi_{p_e}\overline{P}_{{\rm out},12}\;\overline{\epsilon}\bigg[p_2\frac{\lambda_{2{\rm r}}}{\mu_{2{\rm r}}}+1-\frac{\lambda_{2{\rm r}}}{\mu_{2{\rm r}}}\bigg]\\
&\,\,\,\,\,\,\,\,{\rm s.t.}\,\,\,\,0\le \epsilon,p_{1}, p_{2},f_1,f_{2},\rho \le 1\\
&\,\,\,\,\,\,\, \,\,\,\,\,\,\,\,\,\,\,\,\lambda_{\rm p} \leq \mu_{\rm p},\, \lambda_{1{\rm r}}\leq \mu_{1{\rm r}},\,\lambda_{2{\rm r}}\leq \mu_{2{\rm r}},\,\lambda_1\leq \mu_1
\end{split}
\end{equation}

\subsubsection{The case of noncooperation}

We can obtain another inner bound for the system by setting $f_1$ and $f_2$ to zero, thereby inhibiting the cooperation between the primary and secondary users. This system is obviously inferior to the original system with $f_1, f_2 \in [0,1]$ unless, for some system parameters, the optimal values of $f_1$ and $f_2$ are zero making the systems equivalent. Under the scenario without cooperation, queues $Q_{1{\rm r}}$ and $Q_{2{\rm r}}$ are permanently empty and we are left only with the interacting queues $Q_1$ and $Q_2$. These queues can be analyzed exactly in a straightforward manner \cite{ShafieSultan}, but we omit the details here due to space constraints. Our inner bound on the maximum stable throughput region is given by the {\bf union} of the stability regions obtained by analyzing the aforementioned two dominant systems and the system without cooperation.

\vspace{-0.2 cm}

\subsection{Outer Bound}

We consider here an outer bound for the proposed system. Note that ${\rm Pr}\{Q_2=0,Q_{2{\rm r}}=0\} \leq {\rm Pr}\{Q_2=0\}$ and $p_1{\rm Pr}\{Q_{1{\rm r}}\neq 0\}+{\rm Pr}\{Q_{1{\rm r}}=0\} \leq {\rm Pr}\{Q_{1{\rm r}}\neq 0\}+{\rm Pr}\{Q_{1{\rm r}}=0\}=1$. This means that 
\begin{equation}
\begin{split}
\mu_1 \leq \pi_{p_e}\overline{P}_{{\rm out},11}\bigg[\epsilon+\overline{\epsilon} \delta_1{\rm Pr}\{Q_2=0\}\bigg]
\end{split}
\end{equation}
\begin{equation}
\begin{split}
\mbox{Similarly,}\,\,\,\,\mu_2 \leq \pi_{p_e}\overline{P}_{{\rm out},12}\bigg[\overline{\epsilon}+\epsilon \delta_2{\rm Pr}\{Q_1=0\}\bigg]
\end{split}
\end{equation}
\begin{equation}
\begin{split}
\mu_{1{\rm r}}& \leq \pi_{p_e}\overline{P}^{\left({\rm P}\right)}_{{\rm out},11}\bigg[\epsilon+\overline{\epsilon} \delta^{\left({\rm P}\right)}_1{\rm Pr}\{Q_2=0\}\bigg]\cdot \\& \,\,\,\,\,\,\,\,\,\,\,\,\,\,\,\,\,\,\,\,\,\,\,\,\,\,\,\,\,\,\,\,\,\ \bigg[\overline{p_1}{\rm Pr}\{Q_{1}\neq 0\}+{\rm Pr}\{Q_{1}= 0\}\bigg]
\end{split}
\end{equation}
\begin{equation}
\begin{split}
\mu_{2{\rm r}}& \leq \pi_{p_e}\overline{P}^{\left({\rm P}\right)}_{{\rm out},12}\bigg[\overline{\epsilon}+\epsilon \delta^{\left({\rm P}\right)}_2{\rm Pr}\{Q_1=0\}\bigg]\cdot\\& \,\,\,\,\,\,\,\,\,\,\,\,\,\,\,\,\,\,\,\,\,\,\,\,\,\,\,\,\,\,\,\,\,\ \bigg[\overline{p_2}{\rm Pr}\{Q_{2}\neq 0\}+{\rm Pr}\{Q_{2}= 0\}\bigg]
\end{split}
\end{equation}

\noindent Using these upperbounds, we have only two interacting queues, $Q_1$ and $Q_2$. We obtain the stability region by constructing two dominant systems. In the first, user $s_1$ transmits dummy packets when its queue is empty. Under such assumption, the mean service rate of $Q_2$ becomes

\begin{equation}
\begin{split}
\mu_2&=\pi_{p_e}\overline{P}_{{\rm out},12}\overline{\epsilon} \,\,\,\,\,\,\,\,\,\,\,\,\,\,\,\,\,\,\,\,\,\,\,\,\,\,\,\,\,\,\,\,\,\
\end{split}
\end{equation}
\noindent This means that ${\rm Pr}\{Q_2=0\}=1-\frac{\lambda_2}{\pi_{p_e}\overline{P}_{{\rm out},12}\overline{\epsilon} }$.
\begin{equation}
\begin{split}
\mu_1&=\pi_{p_e}\overline{P}_{{\rm out},11}\bigg[\epsilon+\overline{\epsilon} \delta_1\left(1-\frac{\lambda_2}{\pi_{p_e}\overline{P}_{{\rm out},12}\overline{\epsilon} }\right)\bigg]
\end{split}
\end{equation}

\begin{equation}
\begin{split}
\mu_{1{\rm r}}&=\pi_{p_e}\overline{P}^{\left({\rm P}\right)}_{{\rm out},11}\bigg[\epsilon+\overline{\epsilon}\delta^{\left({\rm P}\right)}_1\left(1-\frac{\lambda_2}{\pi_{p_e}\overline{P}_{{\rm out},12}\overline{\epsilon} }\right)\bigg]\overline{p_1}
\end{split}
\end{equation}

\begin{equation}
\begin{split}
\mu_{2{\rm r}}&=\pi_{p_e}\overline{P}^{\left({\rm P}\right)}_{{\rm out},12}\overline{\epsilon}\bigg[\overline{p_2}\frac{\lambda_2}{\pi_{p_e}\overline{P}_{{\rm out},12}\overline{\epsilon}}+1-\frac{\lambda_2}{\pi_{p_e}\overline{P}_{{\rm out},12}\overline{\epsilon} }\bigg]
\end{split}
\end{equation}

The stability region can be characterized by solving the following optimization problem:

\begin{equation}
\begin{split}
&\max_{\epsilon,p_1,p_2,f_1,f_2,\rho} \lambda_1=\pi_{p_e}\overline{P}_{{\rm out},11}\bigg[\epsilon+\overline{\epsilon} \delta_1\left(1-\frac{\lambda_2}{\pi_{p_e}\overline{P}_{{\rm out},12}\overline{\epsilon} }\right)\bigg]\\
&\,\,\,\,\,\,\,\,{\rm s.t.}\,\,\,\,0\le \epsilon,p_{1}, p_{2},f_1,f_{2},\rho \le 1\\
&\,\,\,\,\,\,\, \,\,\,\,\,\,\,\,\,\,\,\,\lambda_{\rm p} \leq \mu_{\rm p},\, \lambda_{1{\rm r}}\leq \mu_{1{\rm r}},\,\lambda_{2{\rm r}}\leq \mu_{2{\rm r}},\,\lambda_2\leq \mu_2
\end{split}
\end{equation}

\noindent The second dominant system to obtain the outer bound can be derived similarly by considering the system in which $s_2$ transmits dummy packets when its queue is empty. The total stability region is given by the union over the two dominant systems.

\vspace{-0.5 cm}
\section{Random Access}
\label{randomaccess}

Assume that the two secondary users sense the channel over the interval $[0,\tau]$ and then access the channel, if sensed to be free from primary activity, with probabilities $\alpha_1$ and $\alpha_2$, respectively. The ranking regarding primary packet decoding is preserved as the ordered system. Hence, the primary service rate and the arrival rates to the relaying queues are identical to the previous section. The other service rates are given by: 

\begin{equation}
\begin{split}
\mu_1&=\bigg[\overline{{\rm Pr}\{Q_2=0,Q_{2{\rm r}}=0\}}\overline{\alpha_2}+{\rm Pr}\{Q_2=0,Q_{2{\rm r}}=0\}\bigg]\\&\cdot\bigg[p_1{\rm Pr}\{Q_{1{\rm r}}\neq 0\}+{\rm Pr}\{Q_{1{\rm r}}= 0\}\bigg]\pi_{p_e}\overline{P}_{{\rm out},11}\alpha_1
\end{split}
\end{equation}

\begin{equation}
\begin{split}
\mu_2&=\bigg[\overline{{\rm Pr}\{Q_1=0,Q_{1{\rm r}}=0\}}\overline{\alpha_1}+{\rm Pr}\{Q_1=0,Q_{1{\rm r}}=0\}\bigg]\\&\cdot\bigg[p_2{\rm Pr}\{Q_{2{\rm r}}\neq 0\}+{\rm Pr}\{Q_{2{\rm r}}= 0\}\bigg]\pi_{p_e}\overline{P}_{{\rm out},12}\alpha_2
\end{split}
\end{equation}

\begin{equation}
\begin{split}
\mu_{1{\rm r}}&=\bigg[\overline{{\rm Pr}\{Q_2=0,Q_{2{\rm r}}=0\}}\overline{\alpha_2}+{\rm Pr}\{Q_2=0,Q_{2{\rm r}}=0\}\bigg]\\&\cdot\bigg[\overline{p_1}{\rm Pr}\{Q_{1}\neq 0\}+{\rm Pr}\{Q_{1}= 0\}\bigg]\pi_{p_e}\overline{P}^{\left({\rm P}\right)}_{{\rm out},11}\alpha_1
\end{split}
\end{equation}

\begin{equation}
\begin{split}
\mu_{2{\rm r}}&=\bigg[\overline{{\rm Pr}\{Q_1=0,Q_{1{\rm r}}=0\}}\overline{\alpha_1}+{\rm Pr}\{Q_1=0,Q_{1{\rm r}}=0\}\bigg]\\&\cdot\bigg[\overline{p_2}{\rm Pr}\{Q_{2}\neq 0\}+{\rm Pr}\{Q_{2}= 0\}\bigg]\pi_{p_e}\overline{P}^{\left({\rm P}\right)}_{{\rm out},12}\alpha_2
\end{split}
\end{equation}

In order to decouple the queue interaction and obtain an inner bound on the stability region of the system, we construct two dominant systems similar to that of the ordered access case and also investigate the case of no cooperation. Due to space limits, we provide here the analysis of the first dominant system only where dummy packets are sent from $Q_1$ and $Q_{\rm 2{\rm r}}$ when they are empty. The mean service rates are given by:

\begin{equation}
\begin{split}
\mu_{1{\rm r}}=\pi_{p_e}\overline{P}^{\left({\rm P}\right)}_{{\rm out},11}\alpha_1 \overline{\alpha_2}\;\overline{p_1},\,\,\,\mu_2= \pi_{p_e}\overline{P}_{{\rm out},12} \overline{\alpha_1} \alpha_2 p_2
\end{split}
\end{equation}

\begin{equation}
\begin{split}
\mu_1&= \pi_{p_e}\overline{P}_{{\rm out},11}\alpha_1\overline{\alpha_2}\bigg[p_1\frac{\lambda_{1{\rm r}}}{\mu_{1{\rm r}}}+1-\frac{\lambda_{1{\rm r}}}{\mu_{1{\rm r}}}\bigg]
\end{split}
\end{equation}


\begin{equation}
\begin{split}
\mu_{2{\rm r}}&=\pi_{p_e}\overline{P}^{\left({\rm P}\right)}_{{\rm out},12} \overline{\alpha_1} \alpha_2 \bigg[\overline{p_2}\frac{\lambda_2}{\mu_2}+1-\frac{\lambda_2}{\mu_2}\bigg]
\end{split}
\end{equation}

The inner bound on the stability region based on the first dominant system is given by the closure of the rate pairs $(\lambda_1,\lambda_2)$. The optimization problem is given by:
\begin{equation}
\begin{split}
&\max_{\alpha_1,\alpha_2,p_1,p_2,f_1,f_2,\rho} \lambda_1= \pi_{p_e}\overline{P}_{{\rm out},11}\alpha_1\overline{\alpha_2}\bigg[p_1\frac{\lambda_{1{\rm r}}}{\mu_{1{\rm r}}}+1-\frac{\lambda_{1{\rm r}}}{\mu_{1{\rm r}}}\bigg]\\
&\,\,\,\,\,\,\,\,{\rm s.t.}\,\,\,\,0\le \alpha_1,\alpha_2,p_{1}, p_{2},f_1,f_{2},\rho \le 1\\
&\,\,\,\,\,\,\, \,\,\,\,\,\,\,\,\,\,\,\,\lambda_{\rm p} \leq \mu_{\rm p},\, \lambda_{1{\rm r}}\leq \mu_{1{\rm r}},\,\lambda_{2{\rm r}}\leq \mu_{2{\rm r}},\,\lambda_2\leq \mu_2
\end{split}
\end{equation}
\noindent We can also obtain an outer bound using the same strategy used in the previous section for ordered access.
\vspace{-0.2 cm}

\section{Numerical Results}
\label{numericalresults}
\begin{figure}
  \includegraphics[width=0.84 \columnwidth]{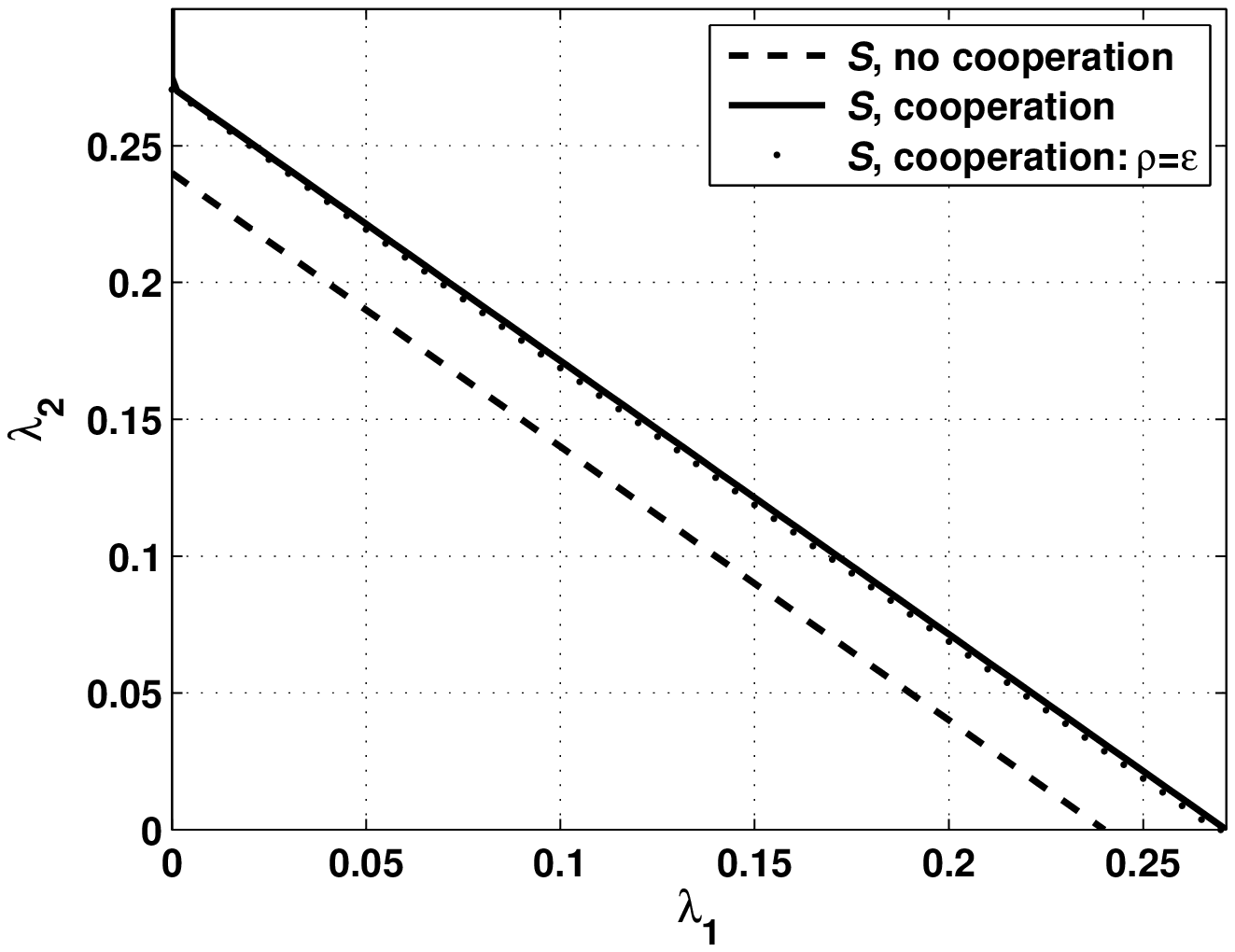}\\
  \caption{Stable throughput region for systems $\mathcal{S}$ with and without cooperation. The parameters used to generate the figure are $1-\frac{\lambda_{\rm p}}{\overline{P}_{\rm out,p}}=0.3$, $\overline{P}_{{\rm out},11}=\overline{P}_{{\rm out},12}=P_{{\rm out},11}^{\left({\rm P}\right)}=0.8$, $P_{{\rm out},12}^{\left({\rm P}\right)}=0.9$, $\delta_1=0.875$, $\delta_2=0.66$, $\delta_1^{\left({\rm P}\right)}=\delta_2^{\left({\rm P}\right)}=0.8$ and $\overline{P}_{{\rm out,p}1}=\overline{P}_{{\rm out,p}2}=0.7$. Also plotted is the stability region for the system with cooperation when $\rho=\epsilon$.}\label{coord_vs_nocoord}
  \end{figure}

We provide here some numerical results. The system with ordered access is called $\mathcal{S}$, whereas the random access system is dubbed $\mathcal{S}^{\left({\rm RA}\right)}$. Fig.\ \ref{coord_vs_nocoord} shows the inner bound for system $\mathcal{S}$, which is the union of the two dominant systems investigated in Section \ref{problemformulation}, together with the case of noncooperation. The stability region is actually specified by the two dominant systems which completely contain the region of stability for the system without cooperation as is clear from the figure. This shows the benefit of cooperation in expanding the secondary maximum stable throughput region. It is important to mention that the noncooperative system has only two interacting queues and its stability region is {\bf exact}, thereby justifying the conclusion on the benefits of cooperation\footnote{The analysis of the case of noncooperation is similar to the two-user ALOHA random access system for which the stability region is exact \cite{rao1988stability}. }. Also plotted in Fig.\ \ref{coord_vs_nocoord} is the system in which the probability $\rho$, related to the order of primary packet acceptance, is equated to $\epsilon$, which is related to the order of secondary spectrum sensing and data transmission. The rationale behind setting  $\rho=\epsilon$ is to reduce the complexity of the problem by decreasing the number of variables. The figure shows that there is little degradation when $\rho=\epsilon$ compared to the case where each parameter is optimized on its own.

\begin{figure}
  \includegraphics[width=0.84 \columnwidth]{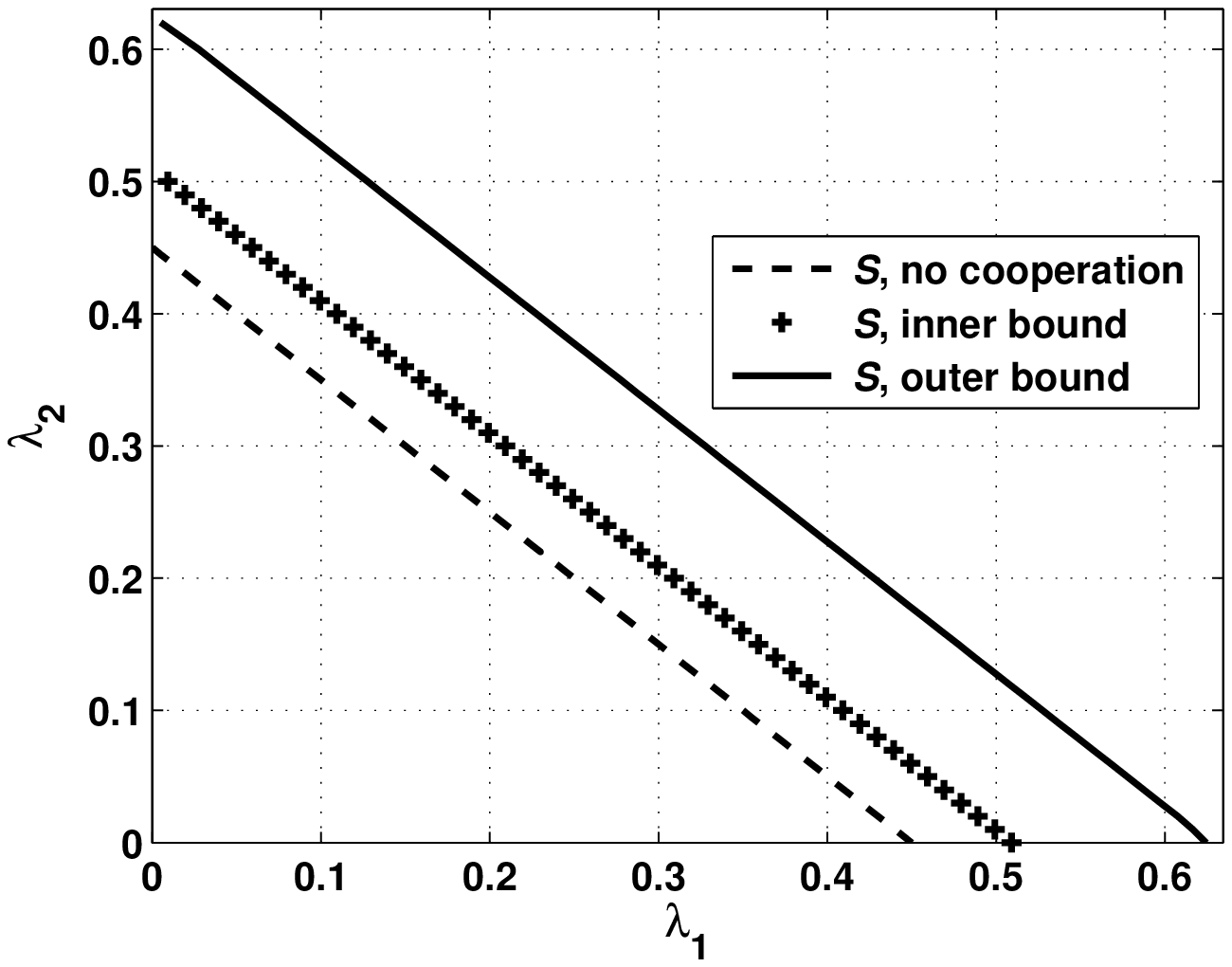}\\
  \caption{Inner and outer bounds on the stability region for system $\mathcal{S}$. The parameters used to generate the figure are $1-\frac{\lambda_{\rm p}}{\overline{P}_{\rm out,p}}=0.5$, $\overline{P}_{{\rm out},11}=\overline{P}_{{\rm out},12}=P_{{\rm out},11}^{\left({\rm P}\right)}=P_{{\rm out},12}^{\left({\rm P}\right)}=0.9$, $\delta_1=\delta_2=0.8$, $\delta_1^{\left({\rm P}\right)}=\delta_2^{\left({\rm P}\right)}=0.9$ and $\overline{P}_{{\rm out,p}1}=\overline{P}_{{\rm out,p}2}=0.85$. The no-cooperation case is also plotted for comparison.}\label{inner_outer}
  \end{figure}

\begin{figure}
  \includegraphics[width=0.85 \columnwidth]{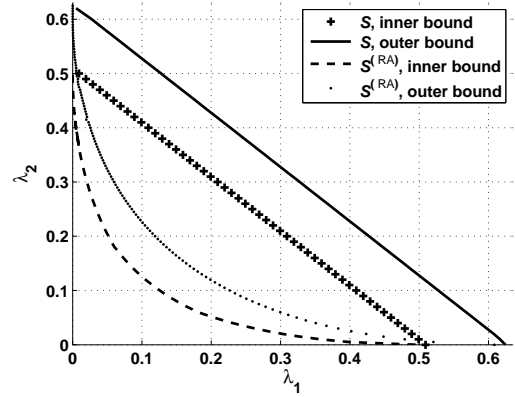}\\
  \caption{Inner and outer bounds on the stability region for the ordered access system $\mathcal{S}$ and random access system $\mathcal{S}^{\left({\rm RA}\right)}$. With the exception of very small arrival rate values, the inner bound on system $\mathcal{S}$ exceeds the outer bound of $\mathcal{S}^{\left({\rm RA}\right)}$.}\label{ordered_vs_random}
  \end{figure}

Fig.\ \ref{inner_outer} shows the inner and outer bounds on the stability of system $\mathcal{S}$ for the parameters given in the caption. The benefit of cooperation is clear from the figure. Fig.\ \ref{ordered_vs_random} provides a comparison between the ordered access and random access schemes. Although we have only inner and outer bounds, it is clear from the figure that for almost all the range of arrival rates, the ordered access scheme is better as its inner bound exceeds the outer bound on the random access scheme.
 
\vspace{-0.4 cm}
\section{Conclusion}
\label{conclusion}
In this work, we studied cognitive relaying in a setting with two secondary users. We showed the possible enhancement of system performance, in terms of an expanded stability region, when the channel access is ordered. Ongoing work includes investigating networks with more than two secondary users. Another research possibility is to account for spectrum sensing errors and cooperation among the cognitive nodes regarding the detection of primary activity. 
\vspace{-0.2 cm}

\bibliographystyle{IEEEtran}
\bibliography{IEEEabrv,shiftting}
\end{document}